# A deep transfer learning network for structural condition identification with limited real-world training data

Nengxin Bao[1, 2#], Tong Zhang[3#], Ruizhi Huang[1, 2], Suryakanta Biswal[4], Jingyong Su[1], Ying Wang[1, 2*]

[1]Harbin Institute of Technology (Shenzhen), Guangdong, China

[2]Guangdong Provincial Key Laboratory of Intelligent and Resilient Structures for Civil Engineering

[3]Peng Cheng Laboratory, Shenzhen, Guangdong, China

[4]University of Surrey, Guildford, UK

## Abstract

Structural condition identification based on monitoring data is important for automatic civil infrastructure asset management. Nevertheless, the monitoring data is almost always insufficient, because the real-time monitoring data of a structure only reflects a limited number of structural conditions, while the number of possible structural conditions is infinite. With insufficient monitoring data, the identification performance may significantly degrade. This study aims to tackle this challenge by proposing a deep transfer learning (TL) approach for structural condition identification. It effectively integrates physics-based and data-driven methods, by generating various training data based on the calibrated finite element (FE) model, pretraining a deep learning (DL) network, and transferring its embedded knowledge to the real monitoring/testing domain. Its performance is demonstrated in a challenging case, vibration-based condition identification of steel frame structures with bolted connection damage. Firstly, disparate subsets of test data are used as training data, and the identification accuracy on the whole data set is evaluated. The results demonstrate that the proposed approach can achieve high identification accuracy with limited types of training data, with the identification accuracy increasing up to 8.57%. Secondly, numerical simulation data are used as training data, and then different TL strategies and different DL architectures are compared on the performance of structural condition identification. The results show that even though the training data are from a different domain and with different types of labels, intrinsic physics can be learned through the pretraining process, and the TL results can be clearly improved, with the identification accuracy increasing from 81.8% to 89.1%. The

---

#These authors contributed equally.

*Prof. Ying Wang is the corresponding Author.

   Email address: yingwang@hit.edu.cn (Prof. Ying Wang)

comparative studies show that SHMnet with three convolutional layers stands out as the pretraining DL architecture, with 21.8% and 25.5% higher identification accuracy values over the other two networks, VGGnet-16 and ResNet-18. The findings of this study advance the potential application of the proposed approach towards expert-level condition identification based on limited real-world training data.

# 1. Introduction

Metallic structures are an important structural type for transport infrastructure, energy infrastructure, and prefabricated buildings [1]. As load transferring elements, bolted connections in such structures are vulnerable due to stress concentrations, with localised damage being particularly hard to detect even under close inspection. However, bolts often suffer from local failure due to progressive corrosion when exposed to harsh environments such as moisture or heavy air contaminants [2]. Previous research including [3] showed that loose bolts could lead to complete loosening of the structural joints, which degrades the overall performance of the infrastructure. Therefore, condition identification of bolted connections is essential to ensure the integrity and safety of the structures.

Structural health monitoring (SHM), which aims to assess the structural conditions accurately and timely, has been one of the most active research areas in civil engineering in the last 30 years [4-7]. Many researchers have investigated SHM methods for bolted connections to ensure the safety of structures in service, which can be categorised as local and global methods. Local methods include impedance-based methods [8], methods based on acoustoelastic effects [9], and displacement sensor-based methods [10], and vision-based methods [11]. However, these methods require the sensors to be located near the target element.

Compared with local methods, vibration-based methods have been more common in civil infrastructure, as a network of closely spaced sensors is not necessary [12]. Research on connection conditions has focused on the development of analytical models with the change in rotational stiffness taken as a damage indicator [13, 14]. While most existing studies focus on frame structures, the rotational stiffness of bridge connections shows similar trends (it decreases with the increase in fatigue damage) [15]. A pilot study [16] demonstrated that the decay rate of vibration signals in impact hammer testing, mostly affected by the damping parameters of connections, can be valuable in understanding connection conditions. It provides not only additional features but the possibility of the direct usage of time-domain signals in detecting connection conditions.

Compared to the above-mentioned methods based on the calculation and comparison of damage indicators, model updating [17] can deliver better performance. Simply speaking, model updating is to find the best match of FE-derived parameters to those obtained from the monitoring data, through iteratively changing the physical parameters in the FE model using an optimisation algorithm, which has been widely applied to structural condition identification [18, 19]. For bolted connection damage, a virtual damper concept and a two-stage model updating scheme have been proposed, which achieved time-domain simulation consistency with real test results [20].

It should be noticed that the computational efficiency of model updating methods is usually low. Therefore, recent research attention has been placed on data-driven methods employing DL algorithms, which have shown promising results. Cha et. al. proposed to use convolutional neural networks (CNN) to detect civil infrastructure defects to partially replace human-conducted onsite inspections [21, 22]. The defects extended from concrete or steel cracks [21] to five types of damages—concrete crack, steel corrosion with two levels, bolt corrosion, and steel delamination [22]. The results showed that the proposed approach can be efficient and accurate. Abdeljaber et al. [23] proposed an enhanced CNN-based approach that was able to successfully identify structural damage caused by loose bolts on steel frames by collecting only the datasets of undamaged and fully damaged cases. The authors proposed SHMnet [24] to identify damage as small as a single bolt loosening in a steel frame based on acceleration time history. Using such methods, the features can be generated automatically and may achieve better structural condition identification results than traditional methods, while computational costs can be significantly reduced [25-27].

Nevertheless, the performance of data-driven methods depends on the quality and quantity of training data. The real-time monitoring data of a structure only reflects a limited number of structural conditions, while the number of possible structural conditions considering damage type, location, and severity, is infinite. With insufficient training data, unsupervised learning methods can be used, using only measured acceleration response data obtained from intact or baseline structures as training data [28]. Nevertheless, the unsupervised learning methods may only be able to detect damage but not to quantify the structural condition. This may restrict further development and application of DL methods in civil engineering.

To cope with this problem, it is generally accepted that only experiments or numerical simulations can be used to supplement the training data [4]. Since experiments are normally expensive and time-consuming, they are unable to meet the amount requirement. Therefore, many researchers use the numerical simulation results as the training data [29, 30]. However, the direct usage of the data derived from finite element simulations may be insufficient for multi-class damage identification problems. For example, optimally selected FE-generated data were used to train DL models for assessing the conditions of simple structures based on laboratory test data, while the identification accuracy for a three-classification problem was only 83.3% [30]. This demonstrates that the distributions in the numerical data and real data are often different. Transfer learning (TL), which aims at boosting the performance of a target model with limited training instances by leveraging the obtained knowledge from different but related source domains, has become a research hotspot recently [31]. Unlike traditional machine learning or DL techniques that usually require abundant labeled training resources, TL focuses on transferring existing knowledge embedded in the pretrained DL models to new target domains with insufficient training data or labels [32].

Most previous works acquire datasets from the same structurally realistic experimental model and employ TL mechanisms to train networks with fewer datasets. However, TL from the numerical domain to the real domain is more challenging and requires further research. According to [31], TL approaches can be interpreted from data and model perspectives. The data-based approaches focus on transferring knowledge via the adjustment and transformation of data. Two strategies are typically implemented to reduce the distribution difference between the source domain and target-domain instances, namely, instance weighting and feature transformation. As for the model-based TL methods, which aim at making accurate predictions on the target domain, it is intuitive to directly share and/or control the parameters of the model trained in the source domain. Such parameter sharing strategy is widely employed in model-based approaches [33]. Gardner et al. [34] used three domain adaptation techniques for damage case classification of structures with either similar or different topologies. Lin et al. [35] proposed a cross-domain structural damage detection method based on deep TL, which generated features insensitive to differences between the model and actual structures by using data from both. Numerical and laboratory results show that the accuracy of the proposed method improved from 83.17% to 93.30%, compared with that of traditional CNN.

Synthesizing the reviewed literature, it can be concluded that the investigation of deep TL is critical for practical SHM applications. Through a challenging case on bolt connection condition identification, this study aims to provide a general structural condition identification paradigm to overcome the contradiction between complex physics and insufficient real data. It provides a detailed discussion of the training data preparation, the network pretraining process, and the TL strategies. The proposed method is validated with the experimental results to demonstrate its effectiveness and applicability. To the best of our knowledge, the work to investigate the deep TL strategy from numerical domain to experimental domain is still rare in the SHM field. The rest of the paper is organized as follows. Section 2 introduces the proposed deep TL approach; Section 3 presents the deep TL algorithm implementation; Section 4 provides the results and discussion of the cases for validating the proposed method; and finally, Section 5 presents the conclusions and possibilities for further development and other applications.

## 2. Methodology

2.1. *Overview*

In this study, a deep transfer learning network, e.g., TL-SHMnet, has been proposed as a structural condition identification scheme. In this study, the dimensions and architecture of TL-SHMnet follows SHMnet [24], as shown in Fig. 1. Nevertheless, the proposed approach is expected to be scalable and general. When applying this network to other cases, the overall architecture will keep the same, while the parameters can be optimized/adapted.

This section gives an overview of the proposed deep TL approach for structural condition identification, by using a subset of test data or FE simulation data for training (referred to as cases 1 and 2, respectively, in the next subsection). The flow chart of the proposed method is shown in Fig 2, which consists of three main steps:

1. Based on the updated FE and state-space models, an optimised FE model is obtained, and numerical simulations are performed to generate training data for multiple types of damage. It should be noted that for case 1, this step reduces to gathering subsets of data only.

2. The network parameters of SHMnet are selected according to the characteristics of the numerical simulation data and used as the baseline model for training and optimization. Then, the training data is augmented by introducing Gaussian noise to the data, and the network is pretrained on the augmented training data set.

3. The pretrained model parameters are transferred to the actual test scenarios by using fine-tuning. Three different strategies are considered by retraining different network parts to find the optimal fine-tuning strategy. The proposed method will be tested for damage identification, including real damage scenarios, and the reliability of the proposed method will be evaluated based on the damage identification results.

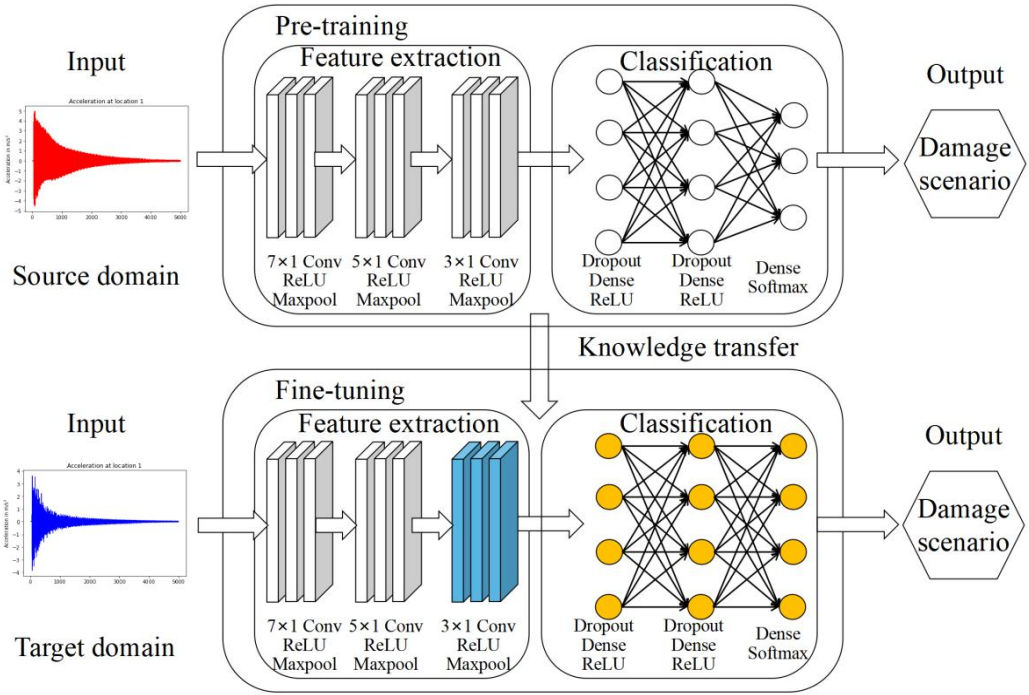

Figure 1. TL-SHMnet: a new deep TL approach

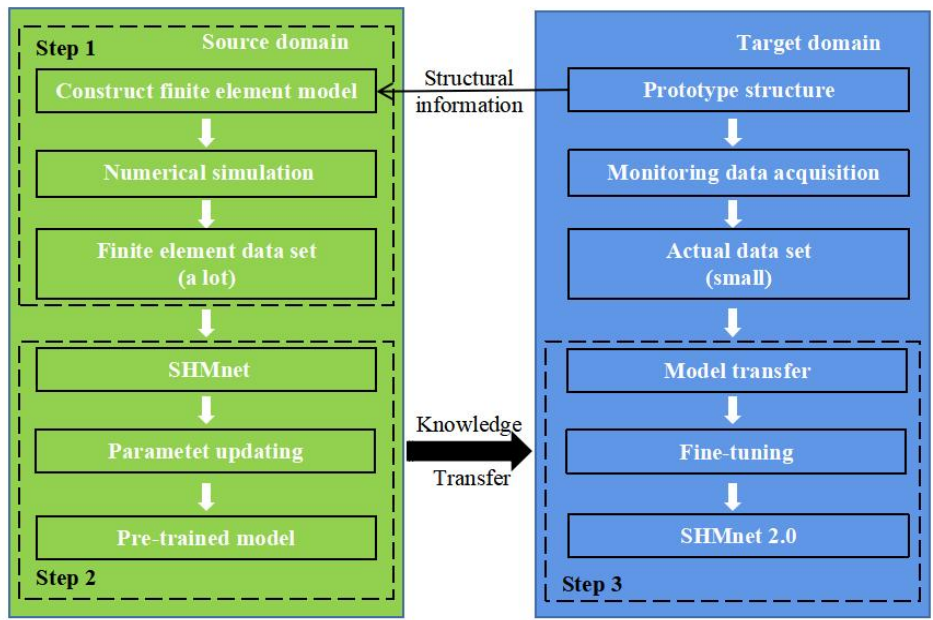

Figure 2. The flowchart of the proposed method

2.2 *Cases considered in this study*

Data-driven structural condition identification methods normally face the challenge that real damage scenarios may not exist in training. As a result, the identification accuracy may be unsatisfactory due to the different distributions between the training and test datasets. Therefore, this study considers two cases to investigate the performance of the proposed method.

Firstly, DL generally shows excellent identification performance by using samples collected from known damage scenarios. Unfortunately, it is difficult to guarantee that the actual damage scenario is within the known range, especially for the damage of bolt connections, because its damage evolution mechanism is still under investigation. In such a case, the direct use of DL method may compromise the accuracy of the identification. To evaluate the performance of the proposed method in this situation, subsets of the damage scenarios in the laboratory test [24] are used as the source domain data to pretrain SHMnet, referred to as **Case 1** thereafter. Subsequently, the network is fine-tuned by using the remaining data sets, which include damage scenarios different from the training scenarios, i.e., loosened bolts at different locations. The main goal is to evaluate the performance of the proposed method in the situation when insufficient test data are available and to provide some suggestions for designing laboratory tests for training data preparation. Based on the results, whether TL can learn general features from limited data will be discussed and the proposed method will be compared with the traditional DL framework.

Secondly, the distribution of numerical simulation data may be different from that of real monitoring data. In extreme cases when the source and target domains are not similar at all, negative transfer situations may even occur. To evaluate the performance of the proposed method in such a situation, this study will use numerical simulation data as the source domain data and laboratory test data as the target domain, referred to as **Case 2** thereafter. In this case, calibrated FE and state-space models are employed to generate time-domain vibration responses of the prototype structure that match the actual test data almost precisely [20]. The results demonstrate the similarity of the data between the source domain and target domain, so negative transfer can be avoided. In this study, systematic numerical simulations are performed, and the results are used as the source domain training data to pretrain the SHMnet. The network is then fine-tuned by using a small amount of actual test data as the target domain. The factors affecting the effectiveness of TL are discussed based on different fine-tuning strategies. The results are compared with the original model without TL to demonstrate the advantages of the proposed method.

To further demonstrate the superior performance of the proposed network, two typical models for each transfer task were compared: VGG-16 [36] and ResNet-18 [37]. To ensure a fair comparison, the feature extractors of the two models were the same and were similarly designed based on the structure of SHMnet feature extractor. Additionally, the TL strategies employed were the same as those discussed earlier.

## 3. Algorithm implementation

3.1 *Training data generation*

3.1.1 *Case 1*

A single-bay, single-storey steel frame was constructed in the laboratory. The details of the steel frame structure and optimally selected sensor locations [38] are shown in Fig 3. Experiments were carried out with ten systematically designed damage scenarios and one intact scenario, as shown in Table 1. Ten repeated impact hammer tests were performed for each scenario. The acceleration responses at each sensor location, together with the impact excitation, were recorded with a sampling frequency of 4096 Hz and a length of 10,000 data points. For each accelerometer, a total of 110 sets of structural responses were recorded. These data sets were not only used to construct SHMnet and to perform the two-stage model updating [20] but also as the training data source for the present study. More details on the experimental works can be found in [24].

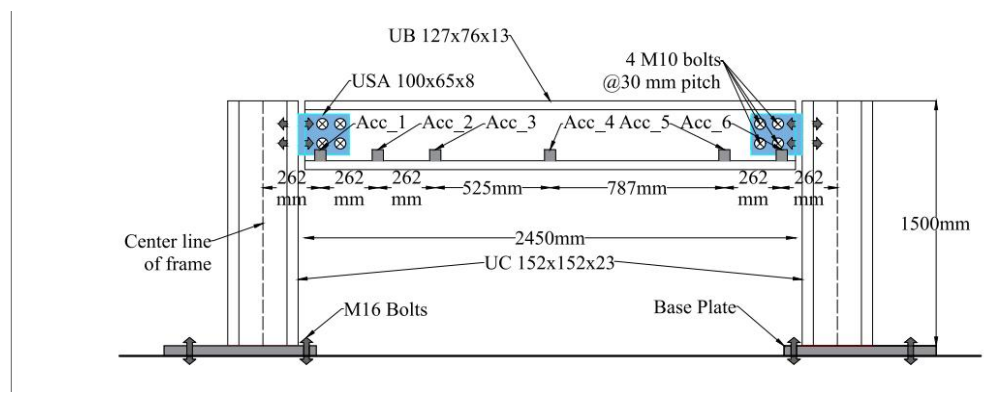

(a)

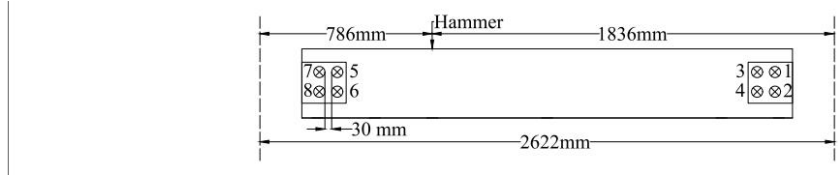

(b)

Figure 3. (a) Portal frame and instrumentation details; (b) Bolt details

Table 1. Designed damage scenarios in [21]

| Detail description | Damage scenario | Category | Bolts loosened* |
|---|---|---|---|
| Intact | 0 | A | All tight |
| One bolt loosened at one end at different location | 1 | B | 1 |
| | 2 | | 3 |
| | 3 | | 2 |
| Two bolts loosened at one end at different location | 4 | C | 1 and 2 |
| | 5 | | 1 and 3 |
| | 6 | | 1 and 4 |
| Two bolts loosened at both ends at different locations | 7 | D | 1, 2, 5 and 6 |
| | 8 | | 1, 3, 5 and 7 |
| Three bolts loosened at one end | 9 | E | 1, 2 and 3 |
| Four bolts loosened at one end | 10 | F | 1, 2, 3 and 4 |

*Numbered according to Fig 3.

As shown in Table 1, ten damage scenarios are considered in addition to the intact scenarios. By categorizing similar damage scenarios into a single group, they can be reduced to five groups, i.e., intact, one bolt loosened, two bolts loosened, two bolts loosened on both sides, three bolts loosened, and four bolts loosened. In this study, four different selections of damage scenarios are considered as shown in Table 2. Specifically, Task 1 includes five damage scenarios, i.e., 0, 1, 4, 9, and 10, each of which represents a typical group. The laboratory test data from the source domain will be used as training data to pretrain SHMnet. The target domain, which consists of the laboratory test data from the remaining damage scenarios, is used as the testing data to evaluate the performance of the proposed method in the identification of different conditions while within the same group. Tasks 2-4 are used to investigate the performance of the proposed method when the training data from all groups are not available. One data set from each damage scenario is used as training data to fine-tune the network, while the remaining five data sets are used as testing data.

In this case, the number of classes in the target domain is more than that in the source domain. Considering the complexity, a new fully connected (FC) layer will be added to the existing FC layer of the pretrained network. In the TL stage, weights in the convolution layers are frozen, while all FC layers are replaced with new FC layers.

Table 2. Task setting of TL

| Description | Source domain | Target domain |
|---|---|---|
| Task 1 | Damage scenario 0, 1, 4, 9, 10 | Damage scenario 2, 3, 5, 6, 7, 8 |
| Task 2 | Damage scenario 0, 1, 9, 10 | Damage scenario 2, 3, 4, 5, 6, 7, 8 |
| Task 3 | Damage scenario 0, 1, 4, 10 | Damage scenario 2, 3, 5, 6, 7, 8, 9 |
| Task 4 | Damage scenario 0, 1, 4, 7 | Damage scenario 2, 3, 5, 6, 8, 9, 10 |

3.1.2 *Case 2*

To train SHMnet with more damage scenarios, there is a clear need to numerically simulate structural behaviour in the time domain. The traditional FE model updating techniques provide a similarity in the frequency domain, and it is shown in [20] that these methods are not sufficient for estimating the responses in the time domain. On the other hand, the traditional time-domain FE model updating techniques need an efficient selection of parameters for predicting the response time histories accurately.

In [20], the authors proposed a new virtual viscous damper (shown in Fig. 4) and a two-stage model updating method to achieve the similarity in the time domain between simulation and test data. Based on the identified virtual damping forces, the acceleration time history estimated by the state-space model almost perfectly agrees with the actual measured acceleration response, and the identified frequency responses successfully pick up the detailed information. This clearly demonstrates that there are similarities between the data from the two domains and the knowledge embedded in the numerical simulation data has the potential to be learned and transferred to condition identification of real experimental/monitoring data. It provides a promising approach to the generation of training data with user-defined scenarios, which breakthroughs the restriction of experimental/monitoring data to limited scenarios. More detailed descriptions of the FE model updating steps and results can be found in [20].

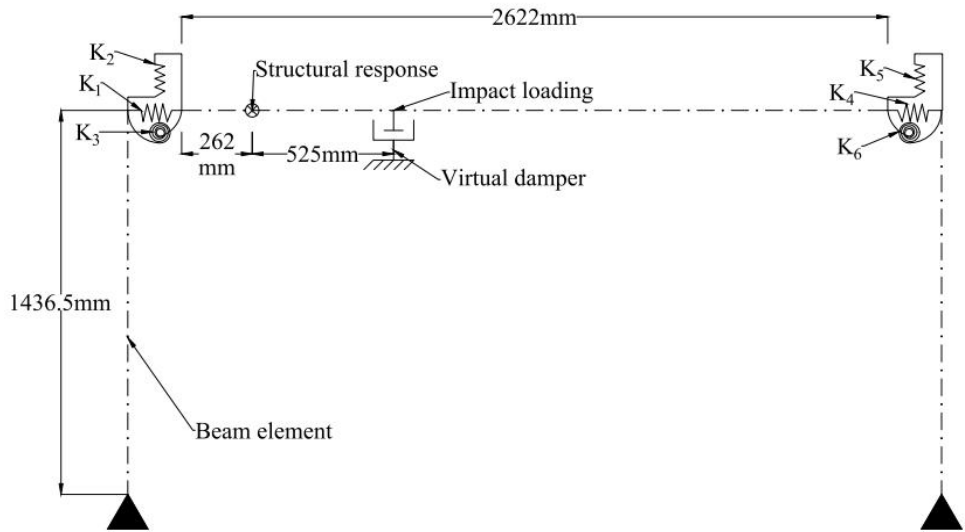

Figure 4. The FE model shows spring and impulse hammer locations

The impact of connection stiffness may be reflected in the natural frequencies and modes of the structure [39]. In this study, the FE model of steel frames uses the element "elasticBeamColumn" to model beams and columns with three springs to represent the complex joint stiffness. As shown in Fig. 4, K1 and K4 act horizontally; K2 and K5 act vertically; while K3 and K6 act rotationally. After accounting for the nonlinear damping effects generated from the structural connections by using a virtual damper, the spring stiffness values have been updated and the estimated accelerations can match the measured acceleration responses almost perfectly. In this part of the study, we generate damage data directly using the updated FE model as the base model.

Once the base model was obtained, numerical simulations using the updated FE model were performed under various conditions by reducing the stiffness of one or more of the springs. Specifically, whole joint damage (K1, K2, and K3), translational spring damage (K1 and K2), and rotational spring damage (K3) were considered for both sides of the frame. Six damage levels were defined by reducing the stiffness of specific components by 2%, 10%, 20%, 50%, and 90%, where a parameter of 0.9 implies a 10% reduction in stiffness, and thus 90% of the original stiffness. The original spring stiffness was already given in [20], and the spring stiffness in damage scenarios are defined according to the percentage changes. In addition to the intact scenario, a total of 36 damage scenarios were simulated, as shown in Table 3.

Table 3. Damage scenarios for FE simulation

|  |  | Description | K1 | K2 | K3 | K4 | K5 | K6 |
|---|---|---|---|---|---|---|---|---|
| Damage scenario 0 |  | Intact | 1.0 | 1.0 | 1.0 | 1.0 | 1.0 | 1.0 |
| Damage scenario 1 | Left | Whole | 0.98 | 0.98 | 0.98 | 1.0 | 1.0 | 1.0 |
| Damage scenario 2 | | | 0.95 | 0.95 | 0.95 | 1.0 | 1.0 | 1.0 |
| Damage scenario 3 | | | 0.9 | 0.9 | 0.9 | 1.0 | 1.0 | 1.0 |
| Damage scenario 4 | | | 0.8 | 0.8 | 0.8 | 1.0 | 1.0 | 1.0 |
| Damage scenario 5 | | | 0.5 | 0.5 | 0.5 | 1.0 | 1.0 | 1.0 |
| Damage scenario 6 | | | 0.1 | 0.1 | 0.1 | 1.0 | 1.0 | 1.0 |
| Damage scenario 7 | Right | Whole | 1.0 | 1.0 | 1.0 | 0.98 | 0.98 | 0.98 |
| Damage scenario 8 | | | 1.0 | 1.0 | 1.0 | 0.95 | 0.95 | 0.95 |
| Damage scenario 9 | | | 1.0 | 1.0 | 1.0 | 0.9 | 0.9 | 0.9 |
| Damage scenario 10 | | | 1.0 | 1.0 | 1.0 | 0.8 | 0.8 | 0.8 |
| Damage scenario 11 | | | 1.0 | 1.0 | 1.0 | 0.5 | 0.5 | 0.5 |
| Damage scenario 12 | | | 1.0 | 1.0 | 1.0 | 0.1 | 0.1 | 0.1 |
| Damage scenario 13 | Left | Translational | 0.98 | 0.98 | 1.0 | 1.0 | 1.0 | 1.0 |
| Damage scenario 14 | | | 0.95 | 0.95 | 1.0 | 1.0 | 1.0 | 1.0 |
| Damage scenario 15 | | | 0.9 | 0.9 | 1.0 | 1.0 | 1.0 | 1.0 |
| Damage scenario 16 | | | 0.8 | 0.8 | 1.0 | 1.0 | 1.0 | 1.0 |
| Damage scenario 17 | | | 0.5 | 0.5 | 1.0 | 1.0 | 1.0 | 1.0 |
| Damage scenario 18 | | | 0.1 | 0.1 | 1.0 | 1.0 | 1.0 | 1.0 |
| Damage scenario 19 | Right | Translational | 1.0 | 1.0 | 1.0 | 0.98 | 0.98 | 1.0 |
| Damage scenario 20 | | | 1.0 | 1.0 | 1.0 | 0.95 | 0.95 | 1.0 |
| Damage scenario 21 | | | 1.0 | 1.0 | 1.0 | 0.9 | 0.9 | 1.0 |
| Damage scenario 22 | | | 1.0 | 1.0 | 1.0 | 0.8 | 0.8 | 1.0 |
| Damage scenario 23 | | | 1.0 | 1.0 | 1.0 | 0.5 | 0.5 | 1.0 |
| Damage scenario 24 | | | 1.0 | 1.0 | 1.0 | 0.1 | 0.1 | 1.0 |
| Damage scenario 25 | Left | Rotational | 1.0 | 1.0 | 0.98 | 1.0 | 1.0 | 1.0 |
| Damage scenario 26 | | | 1.0 | 1.0 | 0.95 | 1.0 | 1.0 | 1.0 |
| Damage scenario 27 | | | 1.0 | 1.0 | 0.9 | 1.0 | 1.0 | 1.0 |
| Damage scenario 28 | | | 1.0 | 1.0 | 0.8 | 1.0 | 1.0 | 1.0 |
| Damage scenario 29 | | | 1.0 | 1.0 | 0.5 | 1.0 | 1.0 | 1.0 |
| Damage scenario 30 | | | 1.0 | 1.0 | 0.1 | 1.0 | 1.0 | 1.0 |
| Damage scenario 31 | Right | Rotational | 1.0 | 1.0 | 1.0 | 1.0 | 1.0 | 0.98 |

| | | | | | | | |
|---|---|---|---|---|---|---|---|
| Damage scenario 32 | | | 1.0 | 1.0 | 1.0 | 1.0 | 1.0 | 0.95 |
| Damage scenario 33 | | | 1.0 | 1.0 | 1.0 | 1.0 | 1.0 | 0.90 |
| Damage scenario 34 | | | 1.0 | 1.0 | 1.0 | 1.0 | 1.0 | 0.8 |
| Damage scenario 35 | | | 1.0 | 1.0 | 1.0 | 1.0 | 1.0 | 0.5 |
| Damage scenario 36 | | | 1.0 | 1.0 | 1.0 | 1.0 | 1.0 | 0.1 |

Each combination of damage was computed ten times with different input impact loadings (previously recorded in the tests [24]), aiming to provide more training data. As shown, if the impact point was well selected, the condition identification results were not sensitive to the sensor location and the time domain structural responses at only one location are needed. Therefore, the effect of sensor location is not discussed in this paper, only the numerical simulation responses of Accelerometer 1, i.e, 262mm from the left end of the beam as shown in Fig 3a, were used.

The sampling frequency was set at 4096 Hz. Each data sample contains 10,000 data points, while they were down sampled to 5000 points each before being fed into the network for training. So the size of the whole dataset is 37x10x5000, which composes the data from the source domain. In addition, the purpose of source domain pretraining is to allow the network to learn about damage in the FE domain thoroughly, which divides the training and test sets in a ratio of 8:2 for pretraining and testing of SHMnet.

3.2 *Network parameters setting and model pretraining*

To identify the subtle damage shown in Table 1, Zhang et al. [24] proposed a DL framework SHMnet, which is a one-dimensional CNN. SHMnet was trained on the time-domain data obtained from experimental works, meaning that time-frequency transformation is not necessary. It achieved 100% identification accuracy for condition identification of the steel frame with bolted connection damage. In comparison, most traditional model-based methods failed this task because the differences between modal parameters under different scenarios are too small. Based on the authors' knowledge, SHMnet is the first open-sourced DL algorithm in the SHM field. It has been used as a benchmark method in recent studies [40]. The details of the network architecture and parameters are shown in Table 4. More details about the algorithm and its application to condition identification of bolted connections can be found at https://github.com/capepoint/SHMnet and Ref

[24].

Table 4. The configuration of the SHMnet architecture

| Layer | Type | Kernel size | Stride | Activation | Input shape | Output shape |
|---|---|---|---|---|---|---|
| 1 | Convolution1d | 7 | 1 | Relu | 1*5000 | 16*4994 |
| 2 | Maxpool1d | 3 | 2 | - | 16*4994 | 16*2496 |
| 3 | Convolution1d | 5 | 1 | Relu | 16*2496 | 64*2492 |
| 4 | Maxpool1d | 3 | 2 | - | 64*2492 | 64*1245 |
| 5 | Convolution1d | 3 | 1 | Relu | 64*1245 | 256*1243 |
| 6 | Maxpool1d | 3 | 2 | - | 256*1243 | 256*621 |
| 7 | Dropout (p=0.5) | - | - | - | 158976 | 158976 |
| 8 | Dense | - | - | Relu | 158976 | 1024 |
| 9 | Dropout (p=0.5) | - | - | - | 1024 | 1024 |
| 10 | Dense | - | - | Relu | 1024 | 1024 |
| 11 | Dense | - | - | - | 1024 | 11 |

The proposed network was implemented on Windows 10 operating system with an Intel Core i9-12900 processor and an NVIDIA GeForce RTX 3090 Ti 24GB GPU. The proposed method is implemented with the DL library Pytorch in the Python IDE Spyder.

Before data from the FE simulation are used as source domain training data for SHMnet, a data augmentation scheme needs to be defined, which is commonly used in DL algorithms to consider various conditions but limited training data. The same scheme as the study [24] is followed by the present study. During the training process, 10% Gaussian noise is applied to the original training data whenever a new batch arrives. After such processing, the input data is converted to a new data set. With the increase in training epochs, the online data augmentation technique can generate increasing volumes of new data, which proves to be an effective way of dealing with the overfitting problem. In addition, this online data augmentation approach does not require much data storage space compared to the offline approach, since the generated data are only used during training. After each training epoch, they will be replaced with another set of newly generated data. It is worth noting that while new data can be created at each epoch, the generated data are highly correlated with the original data, meaning that the intrinsic physics is kept during the process.

3.3 *Fine-tuning*

In this study, the focus will be on the fine-tuning approach for TL. For case 1, subsets of the laboratory test data are used as the source domain, while for case 2, the FE simulation data are selected. These data are used for model pretraining. The target domain data are obtained from laboratory experiments, including the intact scenario and ten damage scenarios, and the test for each scenario is repeated ten times, so the size of the target domain data set is 110x10000. Some labelled data from the target domain are used to assist in training the transfer model for the target task.

For fine-tuning methods, it is important to explore which layers need to be fine-tuned and which layers contain features that can be transferred for structural condition identification problems. In case 2 of this study, the performance under different transfer parameters will be discussed and can be further divided into three standard transfer strategies [33]. The first is to freeze the convolutional layer and fine-tune the FC layer by transferring the weights of the pretrained classifier and adapting it to the target task by fine-tuning it; the second is to fine-tune the convolutional layer and the FC layer on top, and the last is to retain the architecture and weights of the pre-trained SHMnet, transfer the model to the target domain, and fine-tune the whole network model. A detailed image of the transfer strategy is explained in Fig 5.

A series of comparative experiments are used to validate the generalization of model damage identification under different TL strategies. Finally, an optimal TL strategy that balances the calculation speed and accuracy will be obtained.

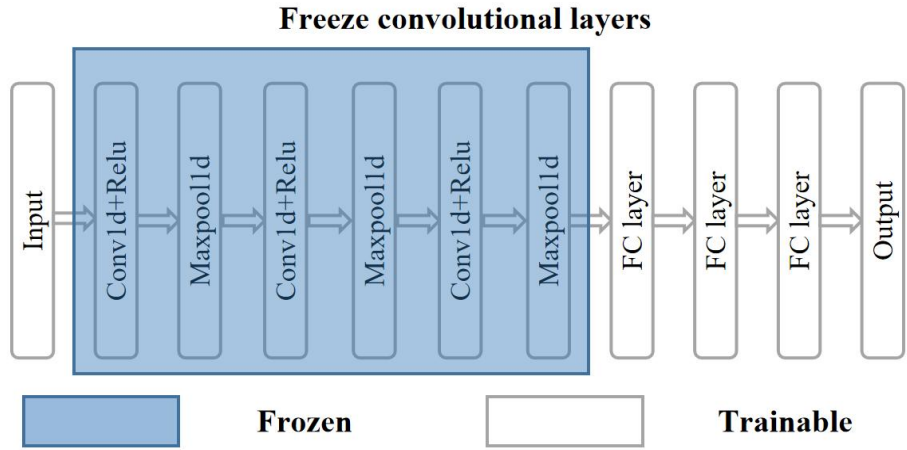

(a) Schematic explanation of Freeze convolutional layers (Strategy 1)

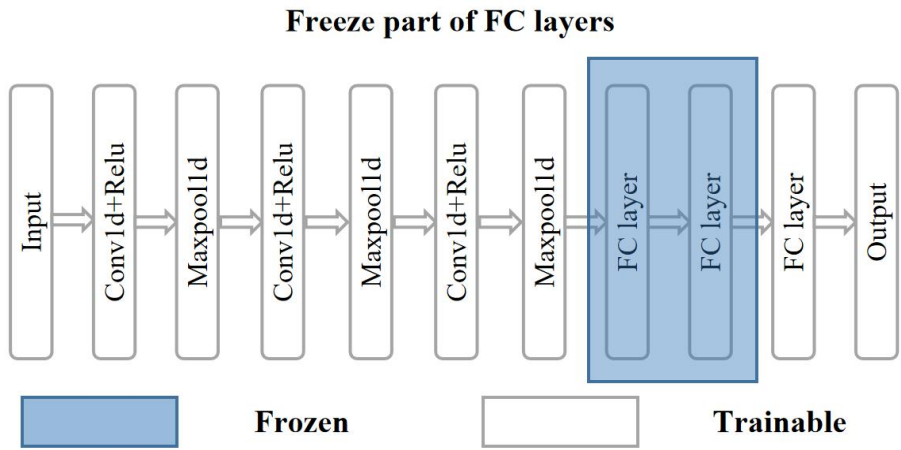

(b) Schematic explanation of Freeze part of FC layer (Strategy 2)

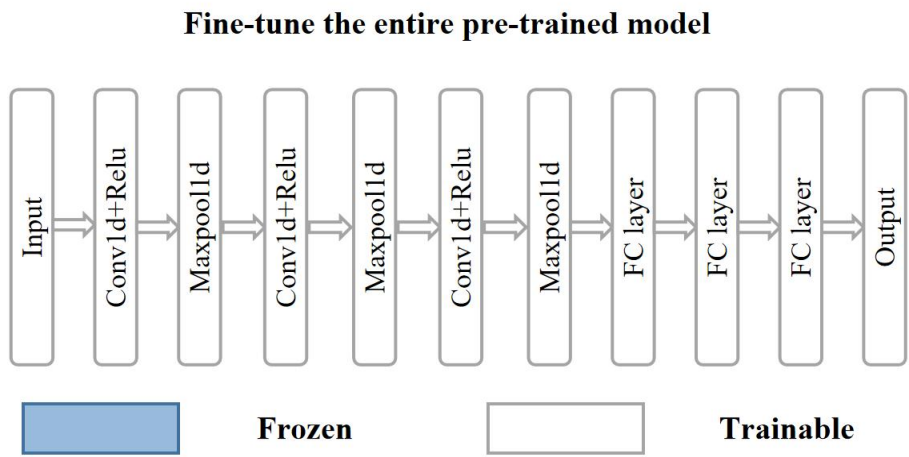

(c) Schematic explanation of fine-tuning the entire pretrained model (Strategy 3)

Figure 5. TL strategies using the pretrained SHMnet model.

# 4. Results and discussions

In this section, two case studies are carried out to verify the effectiveness of the proposed method. The training, fine-tuning, and testing are run on a computational server with configuration as follows: Intel(R) Xeon(R) Gold 6248 CPU @ 2.50GHz, A100 GPU (40GB), 50GB memory.

4.1 *Case 1:TL from laboratory to laboratory*

4.1.1 *Pre-training results*

The SHMnet network is trained using several damage scenarios in the source domain, and each damage scenario is pretrained using five sets of data. The training settings for each task have been provided in Subsection 3.1.1. The testing accuracy is calculated using the remaining five sets of data for each damage scenario. Fig. 6 shows the training/testing results, i.e., the accuracy and loss history curves. (Accuracy = blue line up to the left vertical axis, Loss = red line down the right vertical axis). It can be seen from the curves of each task that convergence can be achieved within 300 epochs. The training time is 778s, 1168s, 958s, and 1029s for Tasks 1, 2, 3, and 4, respectively. For all tasks, the training accuracies are 100%, since the training data are just subsets of the original data from Ref [24], on which SHMnet was trained.

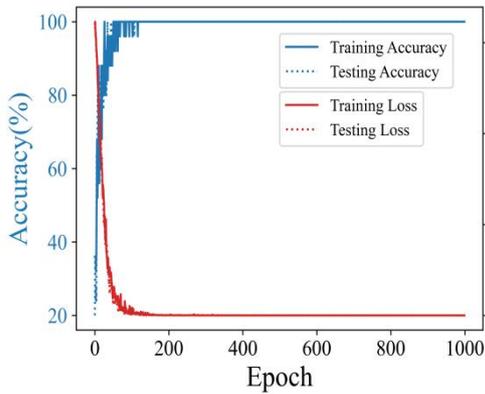 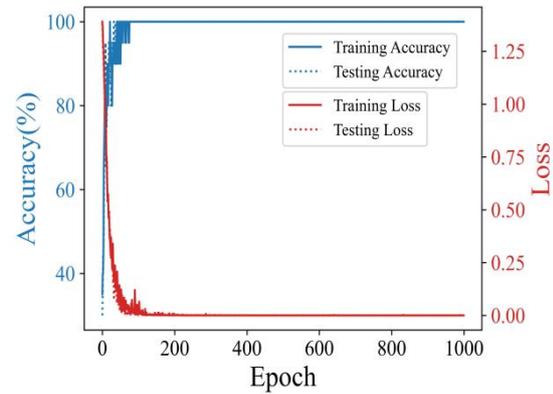

(a) Task 1               (b) Task 2

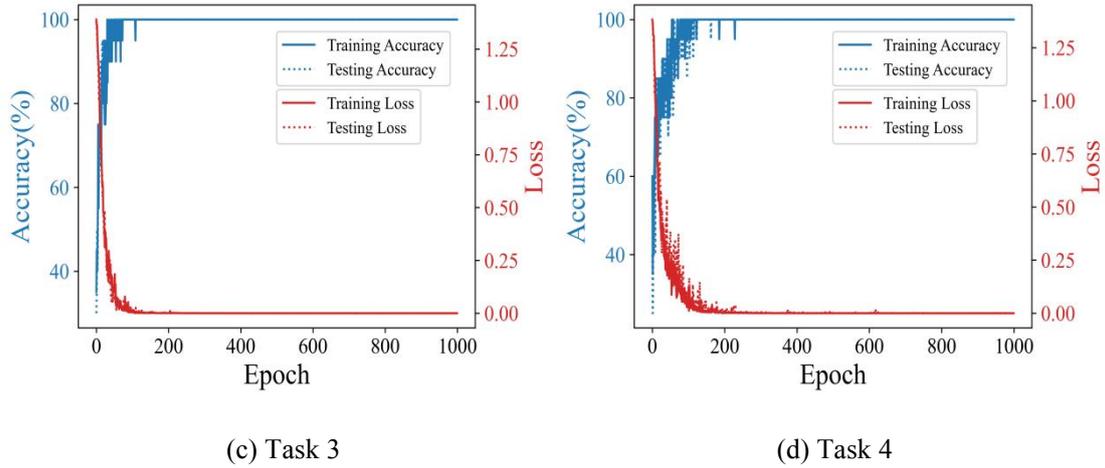

(c) Task 3            (d) Task 4

Figure 6. Training accuracy and loss history of the train datasets: Tasks 1-4

4.1.2 *TL results*

To investigate the generalization ability of the pretrained network, it is then fine-tuned to identify other subsets of untrained damage scenarios. For example, in Task 1, five sets of damage scenarios 0, 1, 4, 7, and 10 were used as the source domain to train the network. The trained network was fine-tuned using just one set of the remaining damage scenarios 2, 3, 5, 6, 8, and 9. After the training was completed for each task, five repeated test results were used as testing data to obtain the damage condition identification accuracy. The TL processes for the other tasks are the same as task 1. Their results are summarized in Table 5 and visualized in Fig 7.

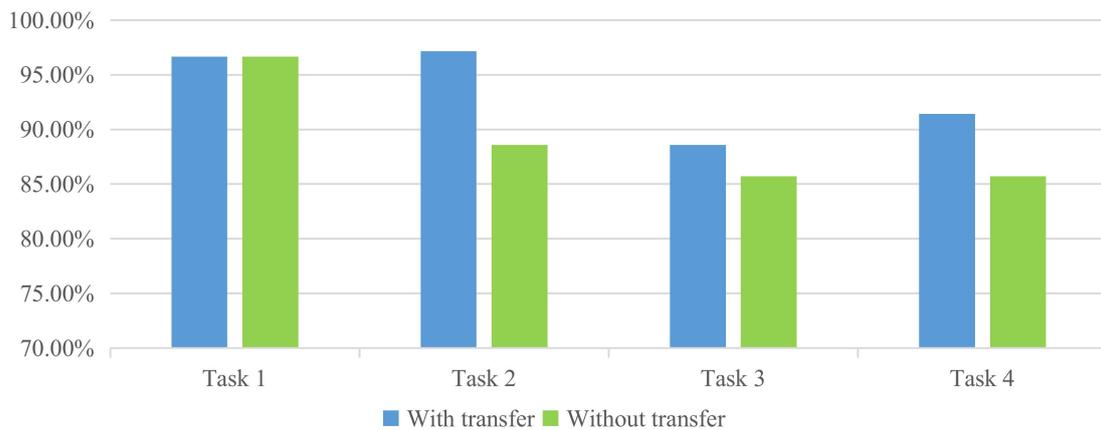

Figure 7. Comparison of condition identification accuracy of each task

Table 5. Comparison of condition identification accuracy and training time of each task

|  | With transfer | | Without transfer | | Accuracy improvement (%) | Time Reduction (s) |
| --- | --- | --- | --- | --- | --- | --- |
|  | Accuracy (%) | Time (s) | Accuracy (%) | Time (s) |  |  |
| Task 1 | 96.67 | 522 | 96.67 | 529 | 0 | 7 |
| Task 2 | 97.14 | 594 | 88.57 | 614 | 8.57 | 20 |
| Task 3 | 88.57 | 596 | 85.71 | 610 | 2.86 | 14 |
| Task 4 | 91.42 | 608 | 85.71 | 617 | 5.71 | 9 |

As can be seen, the accuracy of damage condition identification for tasks 1-4 using the TL method has been improved by 0%, 8.57%, 2.86%, and 5.71%, respectively. For task 1, the lack of significant improvement in damage condition identification accuracy is because the damage subset used for fine-tuning contains all representative damage categories and thus can learn all important features. Therefore, without the pretraining knowledge, the network could identify damage conditions with high accuracy, and this leads to insignificant TL effects. For tasks 2-4, the selected damage subsets are not representative or not fully representative. The TL method allows generic damage features to be extracted, which effectively improves the accuracy of damage condition identification.

Regarding the testing duration, since this method uses a frozen convolutional layer approach, only 54,720 parameters were frozen after the transfer, leaving 163,853,323 parameters still available for fine-tuning. Therefore, little difference in training time between the transfer and without transfer methods has been found.

To analyze the TL effects in more detail, a confusion matrix for damage condition identification was obtained for each task, as shown in Fig. 8. As can be seen in Tasks 1-4, the confusion matrix results show that damage scenarios 2 and 3 are indistinguishable with and without transfer. It means that when the damage is small, the method can correctly detect and quantify the damage, while cannotting identify the damage bolt location.

In Task 2, the confusion matrix shows that the network identifies damage scenario 4 as 7 by prediction without TL. Both damage scenarios 4 and 7 have loose bolts 1 and 2, and the only difference is that the latter also has loose bolts 5 and 6 on the other side. Since the training data for both scenarios were not available for this task, it is reasonable that the pretrained network could not discern these two scenarios. After TL, the structural condition identification accuracy is significantly improved. It demonstrated that fine-tuning could transfer the damage knowledge in the form of parameters even when the initial training data are limited and that only a small amount of data is needed for fine-tuning.

In contrast, although Tasks 3 and 4 presented improvements after TL, it is not as significant as that of Task 2. Two reasons can be attempted to explain these. First, the selection of extreme damage scenarios is important. The comparison between Tasks 2 and 4 shows that extreme damage scenario 10, a damage scenario with four bolts damaged simultaneously, is important for the accuracy of damage identification. The reason behind this is that the extreme damage scenario may contain distinguishable common damage features which can be learned through the fine-tuning process, which lead to significantly improved identification results. Secondly, the selection of similar scenarios may contribute to the improvement of the TL results. For example, Tasks 2 and 4 have very similar damage scenarios 3 and 9, and 4 and 7, respectively. Due to their similarity, the learned features can identify the difference between similar scenarios to improve the identification accuracy after TL. These findings guide our further research on the TL from numerical data to laboratory test data.

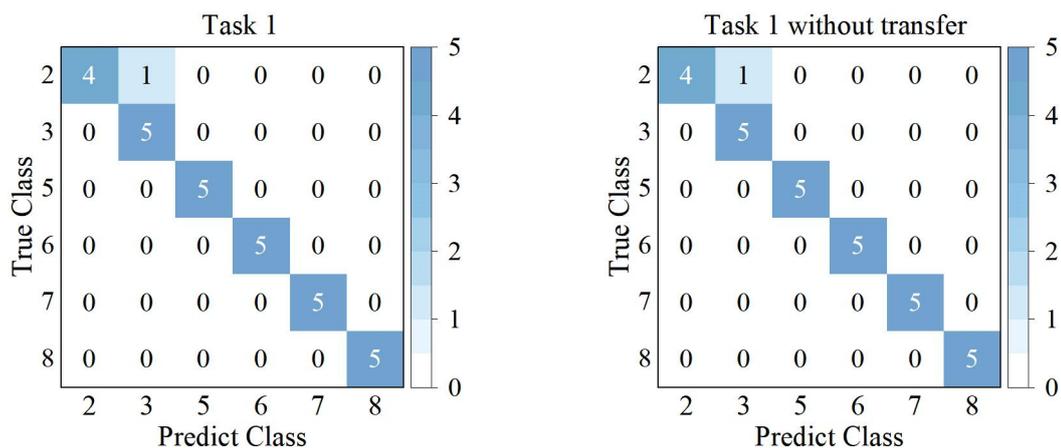

(a) Task 1

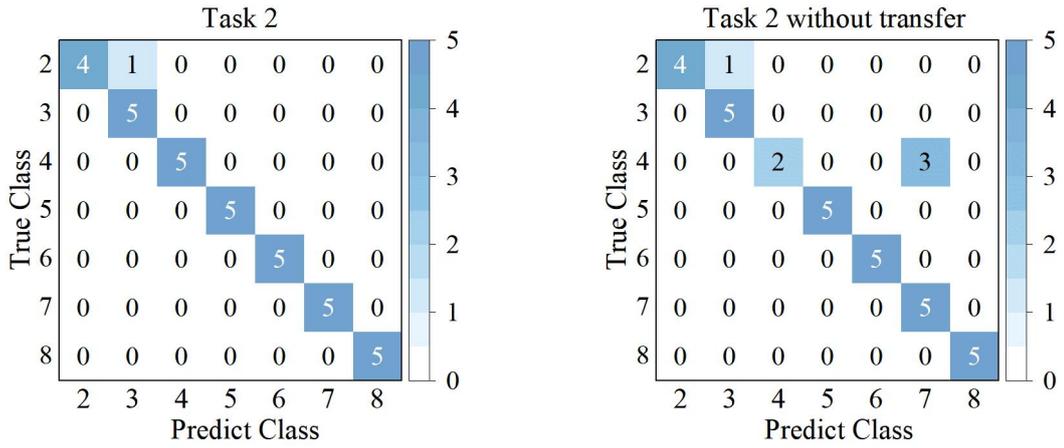

(b) Task 2

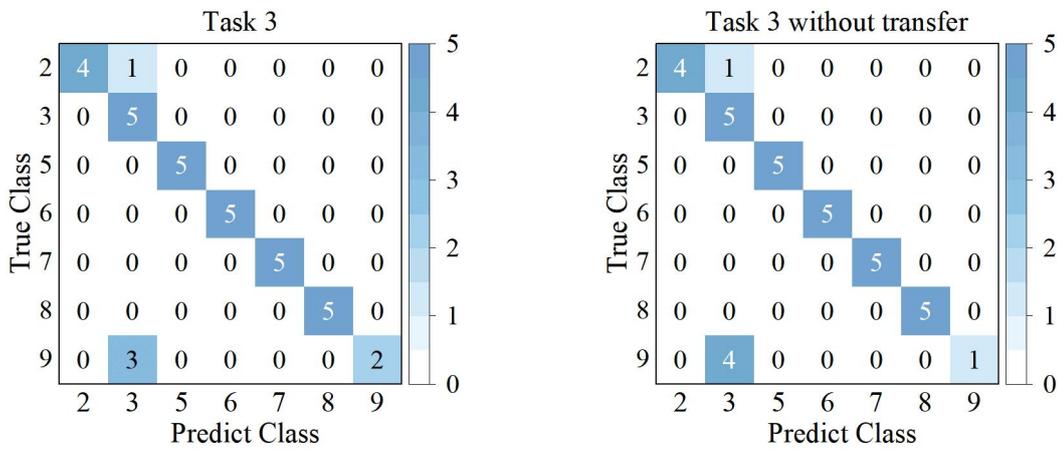

(c) Task 3

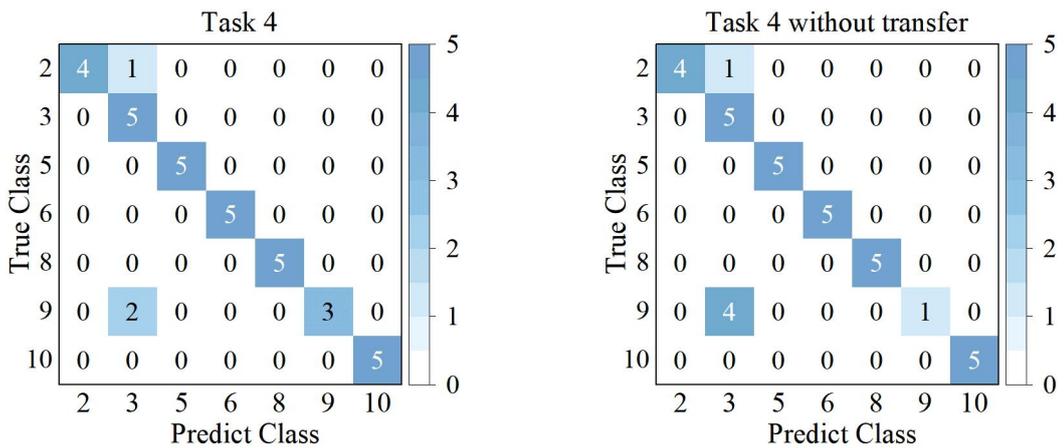

(d) Task 4

Figure 8. Damage identification confusion matrix with and without TL

4.2 *Case 2:TL from FEM to laboratory*

### 4.2.1 *Pre-trained results*

Based on previous research [24], the number of training epochs was set to 1000 and the batch size to 32. For the training process, a variant of Adam's algorithm for stochastic gradient descent was used as the optimization algorithm. In the stage of training, a learning rate of $1\times10^{-4}$ was used to train the network. These hyperparameters were selected and optimized based on previous experience in DL practice. Based on extensive parametric studies, the optimal pretrained network structure is obtained and shown in Table 6.

Numerical simulation results are used to train SHMnet, as described in Section 3.3.2. The pretrained results are shown in Fig. 9. It can be seen that the testing accuracy is around 80%.

Table 6. The configuration of the pretrained model architecture

| Layer | Type | Kernel size | Stride | Padding | Activation | Input shape | Output shape |
|---|---|---|---|---|---|---|---|
| Feature extraction | | | | | | | |
| 1 | Convolution1d | 7 | 1 | 0 | Relu | 1*5000 | 16*4994 |
| 2 | MaxPool1d | 3 | 2 | 0 | - | 16*4994 | 16*2496 |
| 3 | Convolution1d | 5 | 1 | 0 | Relu | 16*2496 | 64*2492 |
| 4 | MaxPool1d | 3 | 2 | 0 | - | 64*2492 | 64*1245 |
| 5 | Convolution1d | 5 | 1 | 0 | Relu | 64*1245 | 256*1245 |
| 6 | MaxPool1d | 3 | 2 | 0 | - | 256*1243 | 256*621 |
| Classification | | | | | | | |
| 7 | Dropout(p=0.5) | - | - | - | - | 158976 | 158976 |
| 8 | Dense | - | - | - | Relu | 158976 | 1024 |
| 9 | Dropout(p=0.5) | - | - | - | - | 1024 | 1024 |
| 10 | Dense | - | - | - | Relu | 1024 | 1024 |
| 11 | Dense | - | - | - | Relu | 1024 | 37 |

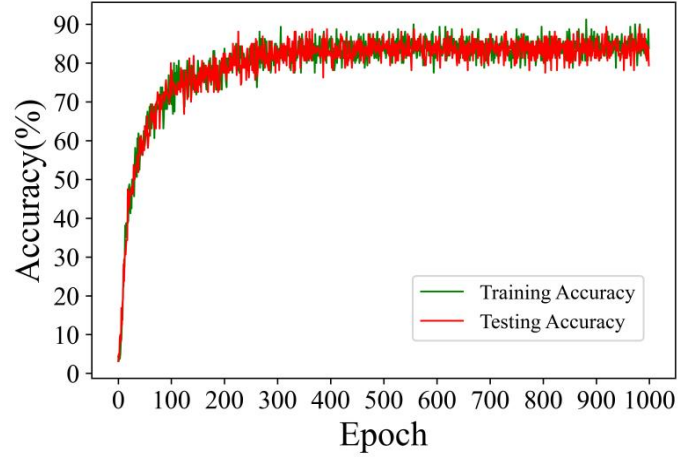

Figure 9. Training and testing accuracy values within the numerical domain

4.2.2 *TL results*

After pretraining on the numerical data, the pretrained network parameters are preserved by freezing different parts of the network structure, and the network is fine-tuned by using laboratory test data. In addition, a transfer-off method is introduced, where SHMnet will be trained on a set of actual damage data directly without a pretraining process. The model feature profiles for each method are summarised in Table 7.

Table 7. Results of different transfer strategies.

| Method | Data structure | | Model parameter | |
| --- | --- | --- | --- | --- |
| | Input size | Training/Testing sample | Trainable | Frozen |
| Transfer off | 5000*1 | 11/55 | 163,908,043 | - |
| Transfer on Strategy 1 | 5000*1 | 11/55 | 163,853,323 | 54,720 |
| Transfer on Strategy 2 | 5000*1 | 11/55 | 65,995 | 163,842,048 |
| Transfer on Strategy 3 | 5000*1 | 11/55 | 163,908,043 | - |

The performance of TL is shown in the confusion matrix in Fig. 10 and summarized in Fig. 11. The accuracy results of each method are 81.8%, 89.1%, 67.3% and 87.3%, respectively. Clearly, Strategy 1 delivers the best performance. The confusion matrix corresponds to a 7.3% improvement in damage identification accuracy before and after transfer. The reason for the better performance of Strategy 1 is that the target experimental domain dataset and the source numerical domain dataset are similar in shallow features, but slightly different in deep features. The shallow layer common to the source task is already extracted from the convolutional layer and damage features from the source task. By keeping the weights of the convolutional kernel constant and fine-tuning the FC layer with the real data, the feature representation can be adapted to the target task and can be a feature representation to the target task, allowing the knowledge of the pretrained model to be reused to reduce the complexity of the model and to reduce the risk of overfitting, thus aiding the real-world damage identification and improving the performance of the model.

For Strategy 2 that only fine-tunes the convolution layers, the model fails to learn the features from real-world data. The basic reason for this phenomenon is that most of the parameters of the network are present in the first two FC layers, which are usually trained on large-scale datasets containing a large amount of knowledge. Therefore, freezing the first two FC layers may inhibit the knowledge transfer of the pretrained model, limiting the expressiveness of the model and thus making it less effective.

While using Strategy 3 that fine-tunes the whole model, the models are overfitted to the limited target domain data. Therefore, Strategy 1 that freezes the parameters of convolution layer retrains the FC layer is both efficient and effective. In such a case, more knowledge learned from the training data can be embedded into the new tasks. It can be concluded that TL can learn standard features from different domains. With limited actual test data for training, the knowledge learned from numerical simulation data can be used to significantly improve the damage condition identification accuracy.

It should be noted that the definitions of damage scenarios in the numerical simulation domain and laboratory test domain are totally different. This is quite often the case in practice, because in simulation, the structure will be simplified to represent the most important static and dynamic mechanism, while in tests, it is more complex. Therefore, the design of numerical simulation scenarios is not necessarily the same as what will happen in real cases. Nevertheless, the results in this study show that even if the numerical simulation scenarios are different from the test scenarios, TL can significantly improve the identification accuracy. This demonstrates the practical effectiveness of TL-SHMnet proposed for structural condition identification.

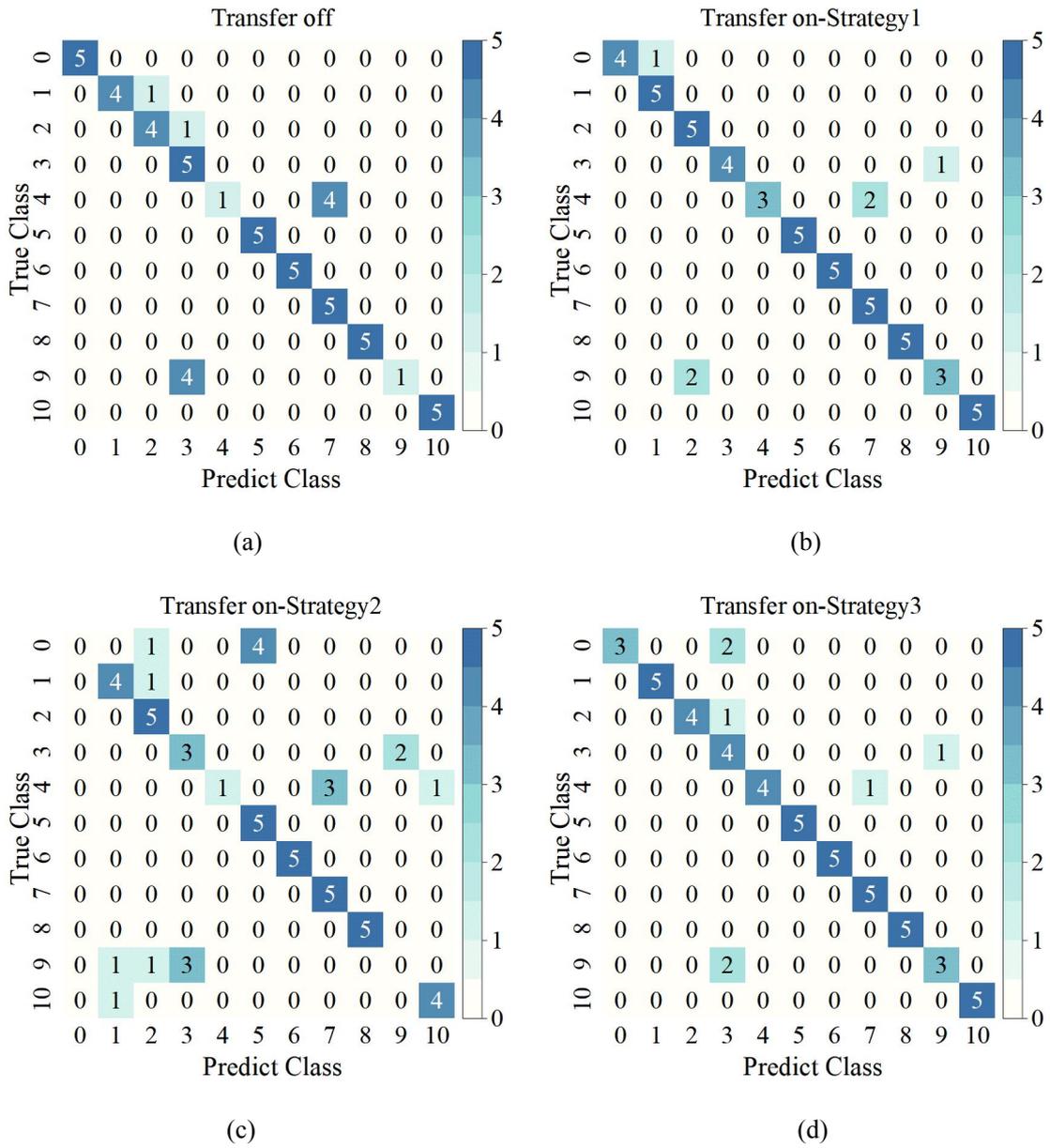

Figure 10. The corresponding confusion matrices for different transfer strategies

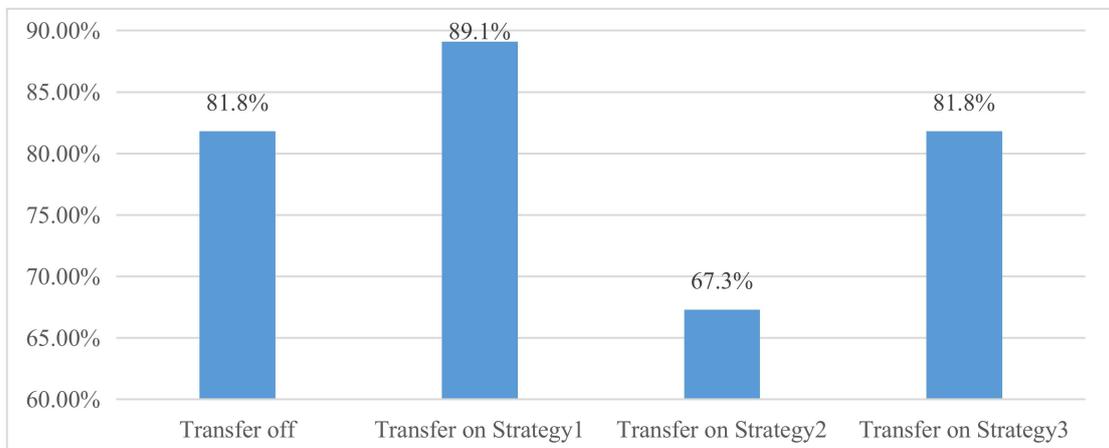

Figure 11. Accuracy of different transfer strategies under real damage scenarios

*4.2.3 Comparison of different convolutional networks*

In order to verify the superior performance of the network used, the classical models VGG-16 and ResNet-18 are also compared, respectively, using the same pretrained and transfer strategies. In traditional CNN architecture, the last set of features is flattened into a high-dimensional feature vector, and each feature is connected to each neuron in the FC layer for classification, which directly leads to a large number of network parameters. Taking the three architectures discussed above, that is, SHMnet, ResNet-18, and VGG-16, as examples, the distributions of parameters between the front frozen convolutional block and the later trainable FC layers are shown in Table 8. It is clear that the majority of parameters are distributed in the FC layer. The convolutional block only occupies a small proportion of parameters, which are responsible for feature learning and extraction. Actually, this part is essential for the condition identification tasks. The accuracy of SHMnet, VGG-16, and ResNet-18 are 89.1%, 67.3%, and 63.6%, respectively, the results are shown in Figure 12.

Based on the results presented in Table 8 and Figure 12, it can be inferred that SHMnet outperforms the other two networks by a significant margin in terms of damage recognition accuracy, with a 21.8% and 25.5% improvement over the other two networks when adopting transfer strategy 1, respectively. Further analysis reveals that despite SHMnet having the highest number of parameters among the three networks, it has fewer convolutional layers due to its architecture. With a small training dataset, an increase in the number of convolutional layers can lead to the overfitting of the network, resulting in a lower identification accuracy in the test set. Meanwhile, the shallow convolutional layers are capable of extracting common damage features in the FE domain, while the deeper convolutional layers extract high-level features that are specific to the corresponding task. As the distribution of damage features in the FE domain may differ from the actual damage distribution, VGG-16 and ResNet-18 fail to perform well with transfer strategy 1, even after freezing the deep convolutional layers, which results in a decrease in their damage recognition accuracy.

Further, the computational durations of SHMnet, VGG-16 and ResNet-18 are compared. For the pre-training stage, the durations are 3.5 hours, 3 hours and 3 hours, respectively. For the fine-tuning stages, the durations are 33 mins, 34 mins and 35 mins, respectively. The results show that the computational efficiency for training different networks is similar. With the training data, the network can be efficiently trained and transferred. Regarding the identification duration, all three networks can deliver the identification results within 1s. Therefore, the proposed approach is promising to fulfil the requirement of real-time structural condition identification in practice.

Overall, the results confirm that TL-SHMnet is a promising approach to accurate and efficient structural condition identification.

Table 8. The main parameter settings of the three models.

| Model | Data structure | | Model parameter | |
|---|---|---|---|---|
| | Input size | Training/Testing sample | Trainable | Frozen |
| SHMnet | 5000*1 | 11/55 | 163,853,323 | 54,720 |
| VGG-16 | 5000*1 | 11/55 | 41,029,131 | 9,831,552 |
| ResNet-18 | 5000*1 | 11/55 | 878,603 | 3,843,648 |

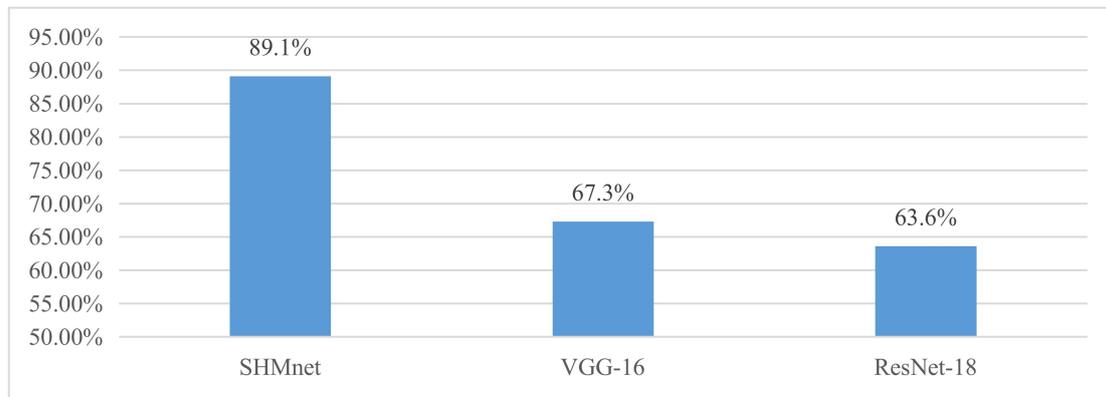

Figure 12. Comparison of accuracy of different networks

4.3 *Discussions*

As a general structural condition identification scheme, the proposed deep TL approach can be extended to practical engineering applications, e.g., structural condition identification of urban bridges. In such a real case, the measurement of structural responses and inputs, e.g., structural displacement/accelerations under random vehicle loads, are needed. The accuracy of real-time data acquisition and processing will be positively related to the accuracy and reliability of the algorithm. More importantly, an accurate and efficient numerical model which can achieve time domain alignment with the real structure such as [20], serving as a structural dynamic digital twin, is needed to generate more training data.

It should be noted that an emerging approach, Generative Adversarial Networks (GANs), can also be used to enrich the training data. GANs have been successfully employed in various domains to address the scarcity of real-world data. The authors proposed to use this approach to complement the limited monitoring data [41]. Ali and Cha successfully applied GANs to the internal damage segmentation using thermography [42]. The use of GANs may largely enhance the efficiency and performance of structural condition identification, and thus is worth further investigation.

# 5. Conclusion

In this paper, a structural condition identification approach based on TL technique, TL-SHMnet, is proposed, which can effectively solve the challenging problems of limited training data and low accuracy in condition identification tasks. The approach is based on previous research outcomes, a DL network SHMnet, and a calibrated FE model for a steel frame with bolted connection damage. The contribution of this study are to demonstrate its effectiveness when a large amount of numerical data, while only a small amount of test data can be used for training, and their definitions of damage are not necessarily the same. Through the case study, the optimal network parameters and transfer strategies are selected and the approach is validated on the laboratory test data of a steel framework. The following conclusions can be drawn:

1. When there are only limited training data, the identification accuracy without TL inevitably degrades. By using the proposed TL approach, the accuracy can be significantly improved. This is correct no matter whether the pretraining data and testing data are from the same domain or not. In this study, for data from different domains, the results increased up to 7.30%.

2. For the case with data from different domains, different transfer strategies may lead to different accuracy results. By freezing the convolutional layers and then fine-tuning the FC layers, the damage condition identification accuracy increased from 81.8 to 89.1%. This may be different from the results in other studies but suitable for SHM tasks with limited real-world data and dissimilar to the numerical domain. Future investigations on large and complex datasets will be conducted to further investigate the reasons to find the optimal design using the TL strategies.

3. Different networks are compared using the same transfer strategy, the results suggest that SHMnet is a promising model for damage recognition tasks in FE simulation, with a 21.8% and 25.5% improvement over the other two networks VGG-16 and ResNet-18.

Overall, the study presents a promising TL approach which can improve structural condition identification accuracy through pretraining and fine-tuning. Future studies can be placed on the application of the proposed approach to real problems, such as urban bridge condition identification. The methods of constructing digital twin models or employing GANs to generate more training data are worthwhile further in-depth investigation. It is expected to provide a new paradigm for structural condition identification and to be of great engineering significance.

## Data availability

The research data will be made available on request by contacting the corresponding author.

## Funding statement

The authors would like to thank Ministry of Science and Technology of the People's Republic of China, for their financial support through National Key Research and Development Program 2022YFB2602103. The computing resources of Pengcheng Laboratory Cloudbrain are used in this research. We acknowledge the support provided by OpenI Community (https://git.openi.org.cn).

# Reference

[1] Wang F and Song G (2020) Bolt-looseness detection by a new percussion-based method using multifractal analysis and gradient boosting decision tree. *Structural Health Monitoring*. 19(6):2023-2032.

[2] Yadavar NSM and Masoud Gi (2019) A Review Paper on Looseness Detection Methods in Bolted Structures. *Latin American Journal of Solids and Structures*. doi:10.1590/1679-78254231.

[3] Mínguez JM and Vogwell J (2006) Effect of torque tightening on the fatigue strength of bolted joints. *Engineering Failure Analysis*, 13(8), 1410-1421.

[4] Farrar CR and Worden K (2012) *Structural health monitoring: a machine learning perspective*. John Wiley & Sons.

[5] Shrestha A, Dang J, Nakajima K, and Wang X (2020) Image processing–based real-time displacement monitoring methods using smart devices. *Structural Control and Health Monitoring*. 27:e2473.

[6] Cadini F, Sbarufatti C, Corbetta M, Cancelliere F, and Giglio M (2019) Particle filtering-based adaptive training of neural networks for real-time structural damage diagnosis and prognosis. *Structural Control and Health Monitoring*. 26:e2451.

[7] Wang Y and Chryssanthopoulos M (2018) Structural condition identification for civil infrastructure: an appraisal based on existing literature reviews. *In EWSHM 2018 Conference proceedings.*

[8] Min J, Park S, Yun CB, Lee CG, and Lee C (2012) Impedance-based structural health monitoring incorporating neural network technique for identification of damage type and severity. *Engineering Structures*, 39(Jun.), 210-220.

[9] Wang T, Song G, Wang Z, and Li Y (2013) Proof-of-concept study of monitoring bolt connection status using a piezoelectric based active sensing method. *Smart Materials and Structures*, 22(8), 087001.

[10] Li J and Hao H (2016) Health monitoring of joint conditions in steel truss bridges with relative displacement sensors. *Measurement*, 360-371.

[11] Yuan C, Chen W, Hao H, and Kong Q (2021) Near real-time bolt-loosening detection using mask and region-based convolutional neural network. *Structural Control and Health Monitoring*, 28:e2741.

[12] Kiranyaz S, Avci O, Abdeljaber O, Ince T, and Inman DJ (2021) 1D convolutional neural networks and applications: a survey. *Mechanical Systems and Signal Processing*, 151(2021).

[13] Yun CB, Yi JH, and Bahng EY (2001) Joint damage assessment of framed structures using a neural network technique. *Engineering Structures*.

[14] Yang JN, Ye X, and Loh CH (2014) Damage identification of bolt connections in a steel frame. *Journal of Structural Engineering*, 140(3), 04013064.

[15] Al-Emrani M (2002) Fatigue in Riveted Railway Bridges-A study of the fatigue performance of riveted stringers and stringer-to-floor-beam connections, PhD Thesis, Chalmers University of Technology.

[16] Wang Y, Ay AM, and Khoo SY (2015) Probability Distribution of Decay Rate: A Novel Damage Identification Method in Time Domain. *The Third Conference on Smart Monitoring, Assessment and Rehabilitation of Civil Structures*.

[17] Friswell M and Mottershead JE (1995) Finite element model updating in structural dynamics (Vol. 38). *Springer Science and Business Media*.

[18] Baisthakur S and Chakraborty A (2020) Modified Hamiltonian Monte-Carlo based Bayesian finite element model updating of a steel truss bridge. *Structural Control and Health Monitoring*. 27:e2556.

[19] Xu Y, Nikitas G, Zhang T, Han Q, Chryssanthopoulos M, Bhattacharya S, and Wang Y (2020) Support condition monitoring of offshore wind turbines using model updating techniques. *Structural Health Monitoring*, 19(4), pp.1017-1031.

[20] Biswal S, Chryssanthopoulos MK, and Wang Y (2021) Condition identification of bolted connections using a virtual viscous damper. *Structural Health Monitoring,* 147592172110092.

[21] Cha, Y.-J., Choi, W. and Büyüköztürk, O. (2017), Deep Learning-Based Crack Damage Detection Using Convolutional Neural Networks. *Computer-Aided Civil and Infrastructure Engineering*, 32: 361-378.

[22] Cha, Y.-J., Choi, W., Suh, G., Mahmoudkhani, S. and Büyüköztürk, O. (2018), Autonomous Structural Visual Inspection Using Region-Based Deep Learning for Detecting Multiple Damage Types. *Computer-Aided Civil and Infrastructure Engineering*, 33: 731-747

[23] Abdeljaber, Osama, Avci, Onur, Kiranyaz, Serkan M, Boashash, Boualem, Sodano, Henry, Inman, and Daniel J (2018) 1-D CNNs for structural damage detection: verification on a structural health monitoring benchmark data. *Neurocomputing*.

[24] Zhang T, Biswal S, and Wang Y (2020) SHMnet: Condition assessment of bolted connection with beyond human-level performance. *Structural Health Monitoring*, 19(4), pp.1188-1201.

[25] Pal J, Sikdar S, and Banerjee S (2022) A deep learning approach for health monitoring of a steel frame structure with bolted connections. *Structural Control and Health Monitoring*. 29(2):e2873.

[26] Ding Z, Li J, and Hao H (2020) Structural damage identification by sparse deep belief network using uncertain and limited data. *Structural Control and Health Monitoring*. 27:e2522.

[27] Ye XW, Jin T, and Yun CB (2019), A review on deep learning-based structural health monitoring of civil infrastructures. *Smart structures and systems*, 24(5), 567-585.

[28] Wang Z and Cha YJ (2020) Unsupervised deep learning approach using a deep auto-encoder with an one-class support vector machine to detect structural damage. *Structural Health Monitoring*, 20(1), 406-425.

[29] Wang Y and Hao H (2015) Damage identification scheme based on compressive

sensing. *Journal of Computing in Civil Engineering,* 29(2).

[30] Seventekidis P, Giagopoulos D, Arailopoulos A, and Markogiannaki O (2020) Structural health monitoring using deep learning with optimal finite element model generated data. *Mechanical Systems and Signal Processing*, 145, 106972.

[31] Zhuang F, Qi Z, Duan K, Xi D, Zhu Y, Zhu H, and He Q (2020) A comprehensive survey on transfer learning. *Proceedings of the IEEE*, 109(1), 43-76.

[32] Weiss K, Khoshgoftaar TM, and Wang D (2016) A survey of transfer learning. *Journal of Big Data,* 3(1).

[33] Han T, Liu C, Yang WG, and Jiang DX (2019) Learning transferable features in deep convolutional neural networks for diagnosing unseen machine conditions. *ISA Transactions*. doi:10.1016/j.isatra.2019.03.017.

[34] Gardner P, Liu X, and Worden K (2020) On the application of domain adaptation in structural health monitoring. *Mechanical Systems and Signal Processing*. doi:10.1016/j.ymssp.2019.106550.

[35] Lin YZ, Nie ZH, and Ma HW (2021) Dynamics-based cross-domain structural damage detection through deep transfer learning. *Computer-Aided Civil and Infrastructure Engineering*. doi:10.1111/MICE.12692.

[36] Biswal S and Wang Y (2019) Optimal Sensor Placement Strategy for the Identification of Local Bolted Connection Failures in Steel Structures. *International Conference on Smart Infrastructure and Construction 2019 (ICSIC)*.

[37] Nikos AS, John SS, and Spilios DF (2019) Vibration-response-only statistical time series structural health monitoring methods: A comprehensive assessment via a scale jacket structure. *Structural Health Monitoring* (3). doi:10.1177/1475921719862487.

[38] Wang F and Song G (2020) Looseness detection in cuplock scaffolds using percussion-based method. *Automation in Construction*, 118.

[39] Simonyan, K., & Zisserman, A. (2014). Very Deep Convolutional Networks for Large-Scale Image Recognition. arXiv:1409.1556

[40] K. He, X. Zhang, S. Ren and J. Sun, (2016) Deep Residual Learning for Image Recognition, *IEEE Conference on Computer Vision and Pattern Recognition (CVPR)*, Las Vegas, NV, USA, pp. 770-778, doi: 10.1109/CVPR.2016.90.

[41] Zhang T and Wang Y (2019) Deep learning algorithms for structural condition identification with limited monitoring data. *International Conference on Smart Infrastructure and Construction 2019 (ICSIC) Driving data-informed decision-making*, (pp. 421-426).

[42] Ali, R., & Cha, Y. J. (2022). Attention-based generative adversarial network with internal damage segmentation using thermography. *Automation in Construction*, 141, 104412.